\documentstyle[aps,amstex,epsfig,prb,multicol]{revtex}
\def\ltsim{\mathop{\raise3pt\hbox{$<$}\llap{\lower3pt\hbox{$\sim   
$}}}}   
\def\gtsim{\mathop{\raise3pt\hbox{$>$}\llap{\lower3pt\hbox{$\sim   
$}}}}   
\newcommand{\be}{\begin{eqnarray}}   
\newcommand{\ee}{\end{eqnarray}}

\newcommand{\gross}{7.9cm}   
   
\newcommand{\gkl}{4.4cm}   
\newcommand{\Gross}{13.9cm}   
\begin{document}   
   
   
\title{Single hole dynamics in the $t$-$J$ model on two- and three-leg ladders}   
\author{Michael Brunner${}^{1,2}$, Sylvain Capponi${}^{1,3}$,  
Fakher F. Assaad${}^{1}$, and Alejandro Muramatsu${}^{1}$}   
\address{   
${}^{1}$Institut f\"ur Theoretische Physik III, Universit\"at Stuttgart,   
Pfaffenwaldring 57, D-70550 Stuttgart,   
Germany\\   
${}^{2}$School of Physics, The University of New South Wales, Sydney, NSW 2052, Australia\\   
${}^{3}$Universit\'e Paul Sabatier, Laboratoire de Physique Quantique,  
118 route de Narbonne, 31062 Toulouse,  
France   
}   
\date{\today}   
\maketitle   
\begin{abstract}  
The dynamics of a single hole in   
the t-J model on  two- (2LL) and three- (3LL) leg ladders is studied    
using a recently developed quantum Monte Carlo algorithm.  
For the 2LL it is shown that in addition to the most   
pronounced features of the spectral function, well  
described by the limit of strong coupling along the rungs,   
a clear shadow band appears in the antibonding channel.   
Moreover, both the bonding band and its shadow have a   
finite quasiparticle (QP) weight in the thermodynamic limit.  
For strong coupling along the rungs of the   
3LL, the low-energy spectrum in the antisymmetric channel is 
similar to a one-dimensional chain, whereas in the two symmetric   
channels it resembles the 2LL.   
The QP weight vanishes in the antisymmetric channel,  
but is finite in the symmetric one.  
\end{abstract}   
%
%
\begin{multicols}{2}   
\narrowtext   
The $n$-leg ladder systems serve  
as a bridge from one-dimensional chains to two-dimensional layered   
cuprates, offering thus a rich playground for studying the interplay of   
charge and spin degrees of freedom in strongly correlated systems  
\cite{dagotto96%
,dagotto99}. 
In particular, two- (2LL) and three- (3LL) leg ladder materials became  
recently available, making experiments possible in such systems
\cite{azuma94}. 
At zero doping, the spin-excitation spectrum   
is gapped in the case of even-leg ladders, whereas it is gapless  
in the case of an odd number of legs, as found by experiments on  
2LL and 3LL \cite{dagotto99,azuma94} and by quantum Monte Carlo (QMC)   
simulations of up to five-leg ladders \cite{frischmuth97}.  
A discussion on  
the impact of this odd-even effect on the charge excitation spectrum as   
obtained by angle resolved photoemission (ARPES) experiments has  
recently started \cite{rice97haas}.  
  
Charge carriers on cuprates can be well described by   
the $t$-$J$ model \cite{zhang88}.  
For the 2LL a considerable amount of work has been done   
by exact diagonalizations (ED)   
\cite{tsunetsugu94
,troyer96%
,haas96},  
density-matrix renormalization group   
(DMRG) \cite{white97}, and strong coupling expansions  
\cite{sushkov99,oitmaa99}, leading to the following picture:  
When the coupling along the rungs $J_{\perp}$ is much larger than along the   
chains ($J_{||}$), the bonding and the antibonding band are split by   
$2t_{\perp}$.   
The bonding band shows a sharp peak at the lower edge of the spectrum.  
Approaching the isotropic case,   
the energy gap between the bonding and antibonding band is reduced,   
and the bands become nearly flat on half of the Brillouin zone.   
Results of exact diagonalizations \cite{troyer96}  
indicate, that spin and charge excitations are coupled to   
each other, leading to the existence of quasiparticles in these systems.  
Like for the 2LL, the strong-coupling limit of the 3LL is a   
promising starting point.   
In this limit, the low-energy excitations have odd parity, and   
can be described   
by an effective one-dimensional (1D) $t$-$J$-model \cite{kagan99},  
which forms a Luttinger liquid in a wide parameter range of $J/t$.  
Results from ED of three-leg ladders  
suggest that the two even channels have the same physics  
as 2LL\cite{rice97haas} up to the isotropic case.  
However, due to the small systems considered, the gap between the even   
and the odd channel is of the same order as the finite-size spin-gap.
Whether a particular channel 
{ will evolve upon doping towards}
the Luttinger-Tomonaga or the Luther-Emery universality class is 
signaled by the absence  
or not of a quasiparticle (QP) weight  
{in the single hole case.}
{This} can only  
be determined by performing a finite-size scaling.
{Whereas a Mott-insulator with branch cuts in the spectral
function (i.e.\ without QP) evolves towards a Luttinger-liquid upon doping,
the Luttinger-liquid parameter reaches the universal value 
$K_\rho=1$ in the limit of zero doping for a spin-gapped system 
\cite{schulz99}.
Then, using a previous analysis of the spectral function for
Luther-Emery systems \cite{voit98} a finite QP weight in the
single hole case results.
}
  
In this paper we study single hole dynamics in the $t$-$J$ model  
on 2LL and 3LL using a   
recently developed QMC algorithm, which has been successfully applied to   
one \cite{brunner99} and two dimensions \cite{brunner00}. The spin  
dynamics is efficiently simulated by a loop-algorithm \cite{evertz93}  
and for each spin configuration, the one-particle Green's function   
is calculated exactly. For details see Ref.~17. 
The QP weight and dispersion 
are  determined from the asymptotics in imaginary time of the Green's
function and  
the full spectral function is obtained with the Maximum Entropy method  
\cite{jarrell96}.  
  
The $t$-$J$ model is given by   
\begin{equation}   
H=-\sum\limits_{\langle i,j\rangle \atop \sigma} t_{ij} 
\tilde c^{\dagger}_{i,\sigma}   
\tilde c^{}_{j,\sigma}   
+\sum\limits_{\langle i,j \rangle} 
J_{ij} 
\bigl( \vec S_i\cdot \vec S_j -\frac 1 4 \tilde n_i \tilde n_j \bigr),   
\label{tJ}   
\end{equation}   
where $\tilde c^{\dagger}_i$ are electron operators restricted to   
the Hilbert space with no double occupancy, $\vec S_i= (1/2)\sum_{\alpha,\beta}   
c^{\dagger}_{i,\alpha}\vec\sigma_{\alpha ,\beta}c^{}_{i,\beta}$, $\tilde n_i=\sum_i   
\tilde c^{\dagger}_i\tilde c^{}_i$ and the   
sum $\langle i,j\rangle$ runs over nearest neighbors only.   
For $i,j$ on a rung we set $J_{ij}=J_{\perp}, t_{ij}=t_{\perp}$, and  
$J_{ij}=J_{||},t_{ij}=t_{||}$ if $i,j$ are on the same leg (   
$t\equiv t_{||}$ is the unit of energy in the following).   
Along the legs of the ladder we use periodic boundary conditions, whereas the boundaries are   
open along the rungs. The reference energy is the Heisenberg groundstate. 
  
We start our discussion with the 2LL in the limit $J_{\perp}\gg J_{||}$.  
At half filling, this leads to the formation   
of singlets on each rung \cite{chubukov89}. 
Perturbative treatments \cite{sushkov99} or series expansions \cite{oitmaa99}  
around this limit show that, like in the free case,  
the hole has a cosine-like dispersion with a splitting of the  
bonding and the antibonding bands by $t_{\perp}$.  
The only difference is a rescaling of the effective hopping to   
$-t_{||}/2$ and a shift of the energies by $J_{\perp}$.   
\begin{figure}[b]
\centering  
\epsfig{file=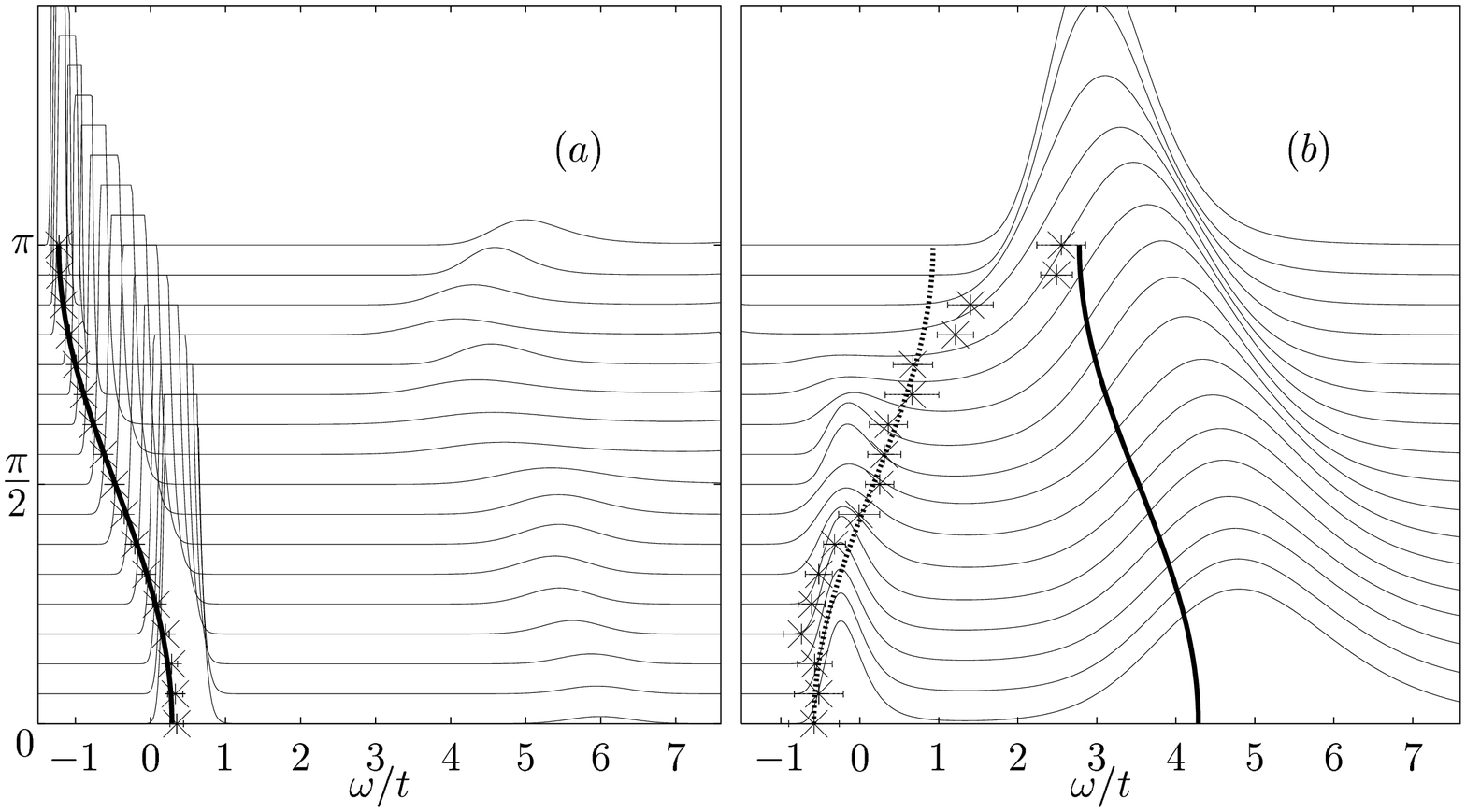,width=\gross}  
\caption{Quasiparticle dispersion  and spectral function for the bonding (a)  
and the antibonding (b) band   
for the 2LL with $J_{\perp}/t_{||}=1.6$, $J_{||}/t_{||}=0.4$,  
and $t_{\perp}/t_{||}=2$. Further details are discussed in the text.  
\label{fig:st2l}}  
\end{figure}  
Figure \ref{fig:st2l} presents our results for a 2LL with 
$2 \times 32$ sites, 
where $J_{||}/t_{||}=0.4$, $J_{\perp}/t_{||}=1.6$, and 
$t_{\perp}/t_{||}=2$. 
{For clarity, the quasiparticle peaks were cut off
at 
a given intensity}\cite{fn1}.
As expected, the major feature in the antibonding channel is   
$4t=2t_{\perp}$ above the bonding band. Both bands can be fitted  
by $(0.75 \cos k_{||} - 0.47) t$ for the bonding and  
$(0.75 \cos k_{||} - 0.47 +4) t$  
in the antibonding band (full lines in Fig.~\ref{fig:st2l}).  
However, we find additionally  
a clear evidence of a shadow band in the antibonding channel. This is   
corroborated by considering apart from the spectral function,   
the QP dispersion directly obtained  
from the imaginary time Green's function ($\times$ with errorbars), and  
by superposing the fitted bonding band with an energy shift of $0.64t$,   
which is approximately the spin gap \cite{barnes93%
,white94}, 
and   
a momentum shift of $\pi$ [dashed line in Fig.~\ref{fig:st2l}(b)], as  
expected for a shadow resulting from the coupling of the bonding   
band to the lowest spin excitations centered at $(\pi,\pi)$ and with an energy 
given by the spin gap.  
The possibility of a shadow band for 2LL was first proposed on the  
basis of ED \cite{haas96}, but due to the small system sizes, no 
quantitative   
assignment was possible. In principle, also a shadow of the antibonding band 
should be expected 
in the bonding channel. However, since the spectral weight  
in the antibonding channel is split between the original band and the  
shadow of the bonding band, resulting in rather broad features in comparison 
to the bonding band, less well defined shadows should be expected in this case.
Only a 
weak structure is observed in Fig.\ \ref{fig:st2l}(a) at around the energy  
of the antibonding band, but it is difficult to assign a dispersion to it. 
 
\begin{figure}[tb]  
\centering  
\epsfig{file=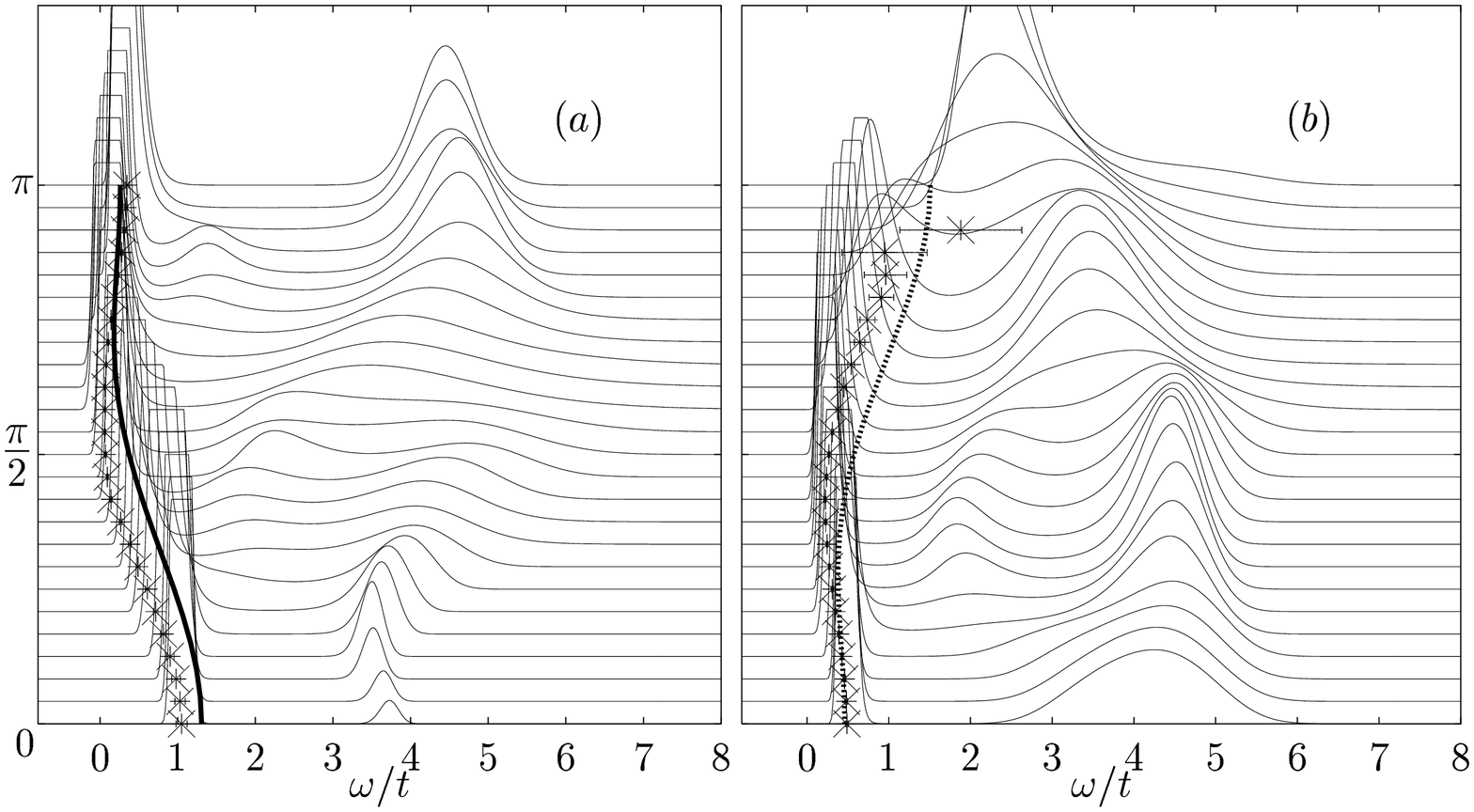,width=\gross}  
\caption{Quasiparticle dispersion for the bonding (a) and antibonding 
(b) band
for the isotropic two-leg ladder with $J/t=1$.  
Further details are discussed in the text.  
\label{fig:i2l}}  
\end{figure}

Figure \ref{fig:i2l} presents the isotropic case  
$J_{||}=J_{\perp}=t_{||}=t_{\perp}$ for the 2LL ($2 \times 48$ sites). 
As before, expansions  
around the strong coupling limit \cite{sushkov99,oitmaa99}  
describe accurately the bonding band obtained from our simulations.  
The high-energy excitations show no clear dispersion, and are about $5t$   
higher in energy in both bands. Again, the antibonding band can be   
interpreted as a shadow of the bonding band. However, the energy difference  
between the bonding band and its shadow ($\Delta=0.17t\pm 0.02t$) is   
now smaller   
than the value of the spin gap ($\sim J/2$) \cite{white94}, which is only an  
upper bound that is reached when the bonding and antibonding bands are well  
separated as in the strong coupling case.  
The lowest states for the hole in both the bonding and its shadow are now   
closer to $k_{||}=\pi/2$ with a band-width $\sim 0.5t$, 
in rough agreement with  
ARPES results on Sr$_{14}$Cu$_{24}$O$_{41}$ \cite{takahashi97} in spite of the
somewhat large value of $J$. 
The bands  
observed in ARPES are symmetric around $k_{||}=\pi/2$, a feature that can be 
reproduced by 
superposing the bonding band and its shadow in Fig.\ \ref{fig:i2l}.  
The very flat portions around $k_{||}=\pi$ ($k_{||}=0$) for the bonding (shadow)
are reminiscent of the flat dispersion around ($\pi,0$) or $(0,\pi)$  
in two dimensions,  
as one approaches these points from $(0,0)$ or $(\pi,\pi)$ respectively,
as already suggested based on a reduced basis approximation \cite{martins99}. 

\begin{figure}[tb] 
\centering 
\epsfig{file=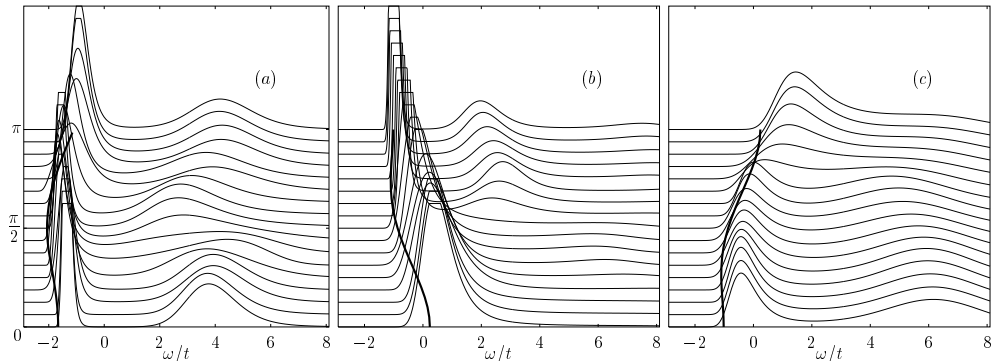,width=\Gross} 
\caption{Spectral function  
for the 3LL with $J_{\perp}/t_{||}=1.6$, $J_{||}/t_{||}=0.4$, 
and $t_{\perp}/t_{||}=2$ in the antisymmetric (a), the symmetric bonding (b) 
and the antibonding (c) band. Further details are discussed in the text. 
\label{fig:st3l}} 
\end{figure}

We consider next the 3LL and start with the strong coupling limit, where  
the low energy behavior was shown to correspond to an effective   
one-dimensional $t$-$J$-chain in the antisymmetric channel \cite{kagan99},  
whereas ED \cite{rice97haas} indicate that the two symmetric channels have a   
finite energy gap  to the antisymmetric one. The observed gap is however  
quite small $\sim 0.3J=0.15t$, and therefore finite-size effects cannot  
be excluded. Results from DMRG \cite{white97,white98} show, that at low   
doping, the holes are at the two outer legs of the 3LL confirming the   
picture derived from ED. Figure \ref{fig:st3l} shows the spectral function  
for a 3LL with $3 \times 32$ sites and the same parameters as for the 2LL.  
As expected, the lowest state belongs to the antisymmetric channel. For  
$J_\perp \gg J_\parallel$, the effective parameters of the corresponding  
$t$-$J$ chain are given by   
$t_{eff} = 3 t_\parallel \left(J_\perp +  
\sqrt{J_\perp^2 + 8 t_\perp^2} \right)/  
8 \sqrt{J_\perp^2 + 8 t_\perp^2}$ and $J_{eff} = J_\parallel$ \cite{kagan99}.  
Using these parameters, and a model with free spinons and holons  
\cite{suzuura97,brunner99}, a reasonably good description of the {lower edge of the } 
antisymmetric band is obtained (full line in Fig.~\ref{fig:st3l}(a) 
corresponds to the minimal energy given by the convolution of spinons and holons). 
Even more interesting are the results obtained 
for the two symmetric bands. They are well separated 
from the antisymmetric one in energy, and can be   
very well fitted by the bonding
\end{multicols} 
\widetext 
\begin{figure}[tb]  
\centering
\epsfig{file=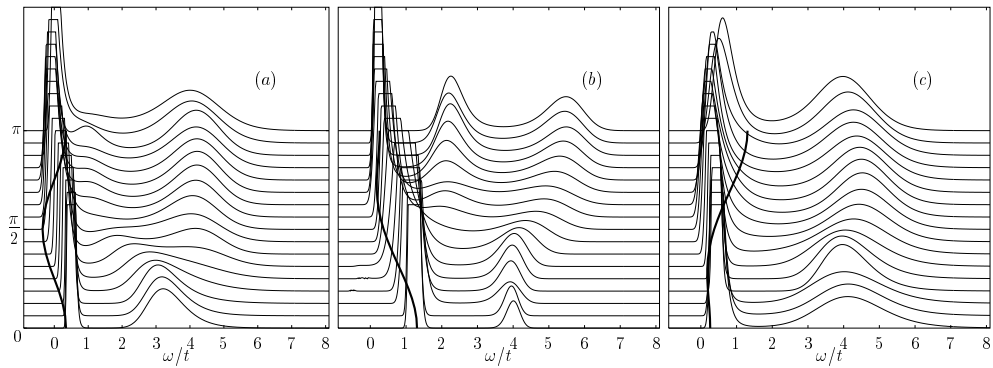,width=\Gross}  
\caption{Spectral function  
for the 3LL  
with $J/t=1$.  
Further details are discussed in the text.  
\label{fig:i3l}}  
\end{figure}  
\begin{multicols}{2} 
\narrowtext  
\noindent 
 
band dispersion for the isotropic   
2LL from perturbation theory \cite{sushkov99}, with $J/t=0.4$, and its shadow  
[full lines in Fig.\ \ref{fig:st3l}(b) and (c)].  
The two bands with even parity are connected by spin excitations with even parity, which are gapless   
\cite{rice97haas,frischmuth96}, such that in this case no shift between them   
appears.  
The description of the 
3LL system at the isotropic point ({$J=t, t_{eff}=0.5,$}$ 3 \times 32$ sites, Fig.\ \ref{fig:i3l})  
on the other hand, is less satisfactory when 
the strong coupling limit is used. All three   
bands have their minima around $3\pi/4$, and contrary to   
ED \cite{rice97haas}, we do not find a significant   
energy gap between all three bands. 
{The shape of
the lower edge  of the antisymmetric channel - which is determined by the holon
dispersion relation  for this value of $J_{eff}/t_{eff} = 2$ -  differs
considerably from that of the  Bethe-Ansatz holon dispersion at the supersymmetric
point $J/t=2$ \cite{sarkar90}.  }

\begin{figure}[tb]   
\centering   
\epsfig{file=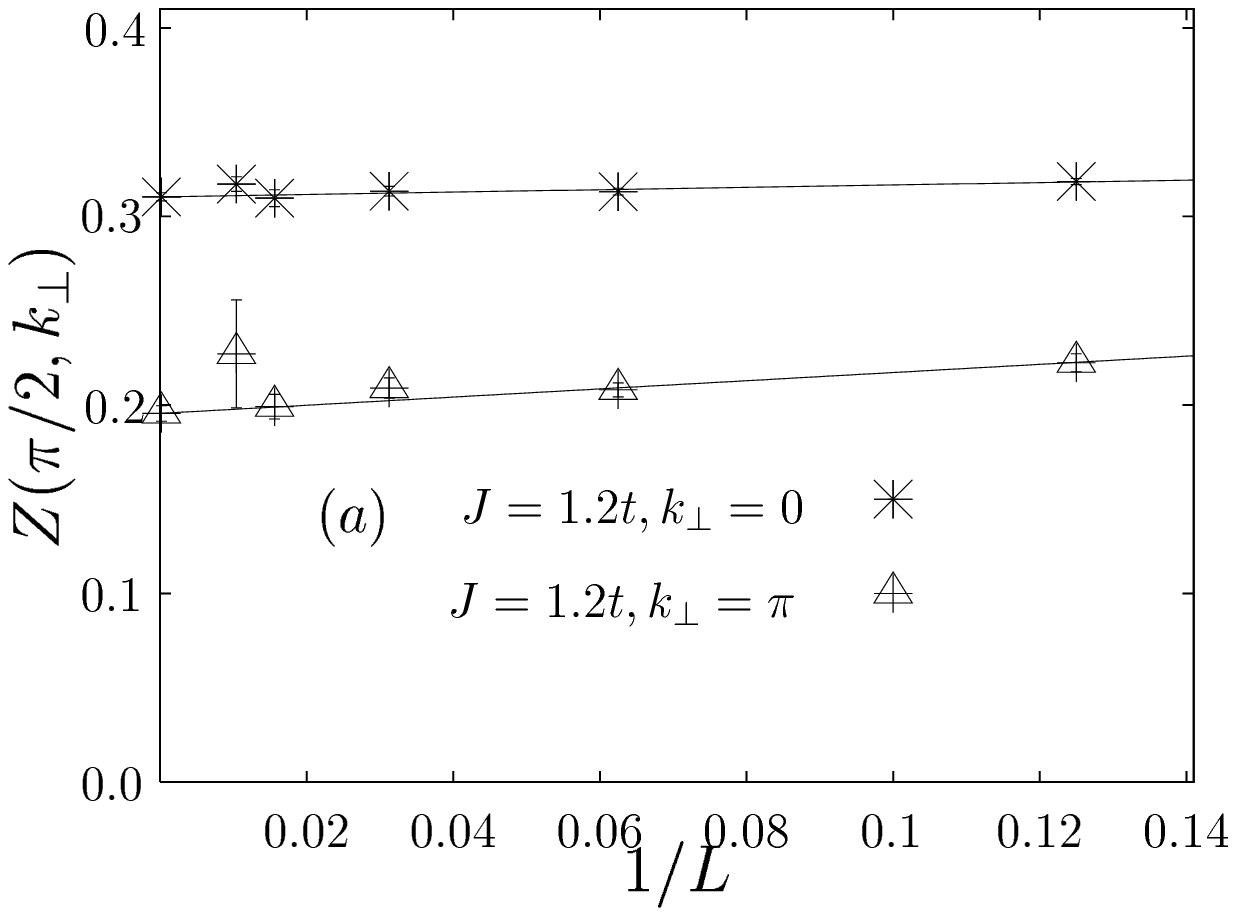,width=\gkl}\epsfig{file=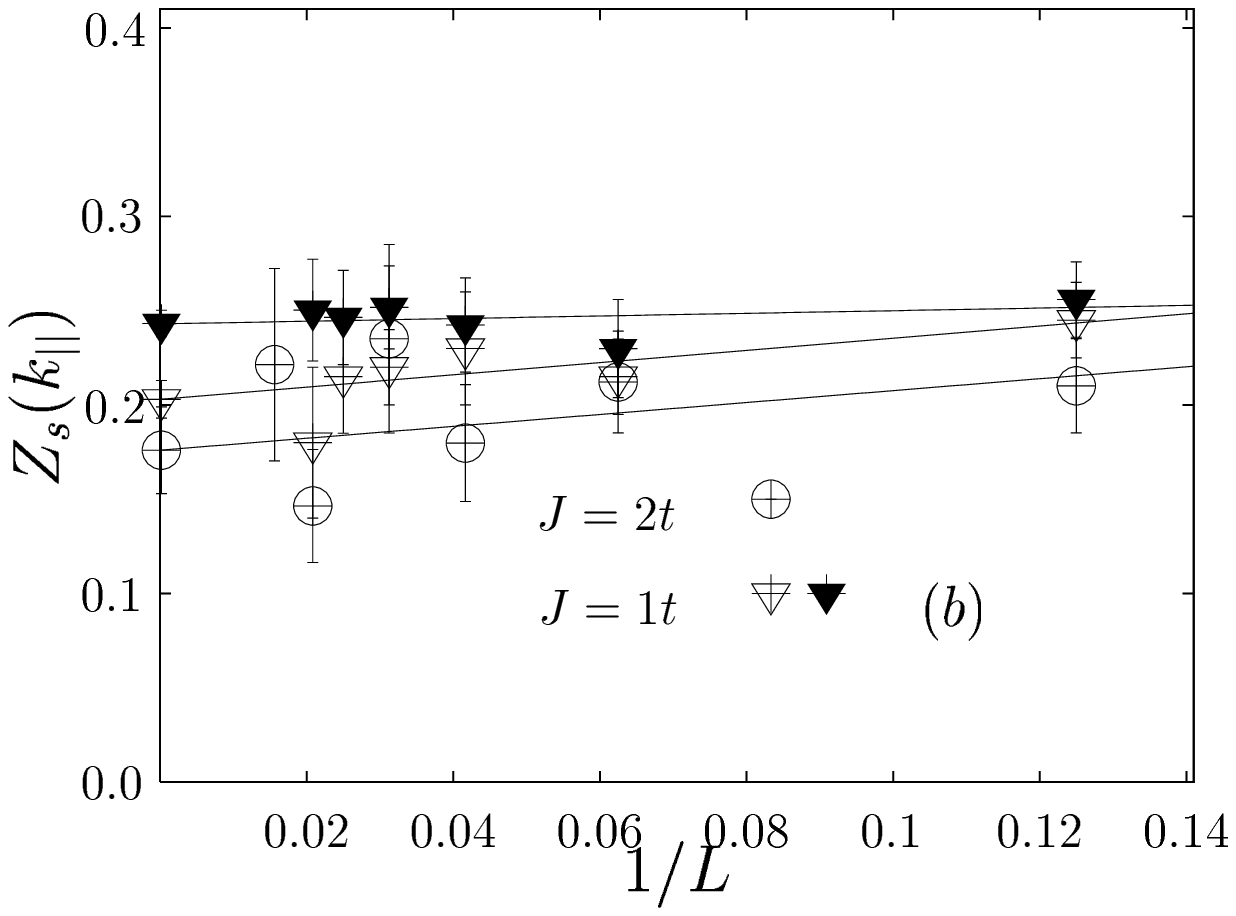,width=\gkl}\\ 
\epsfig{file=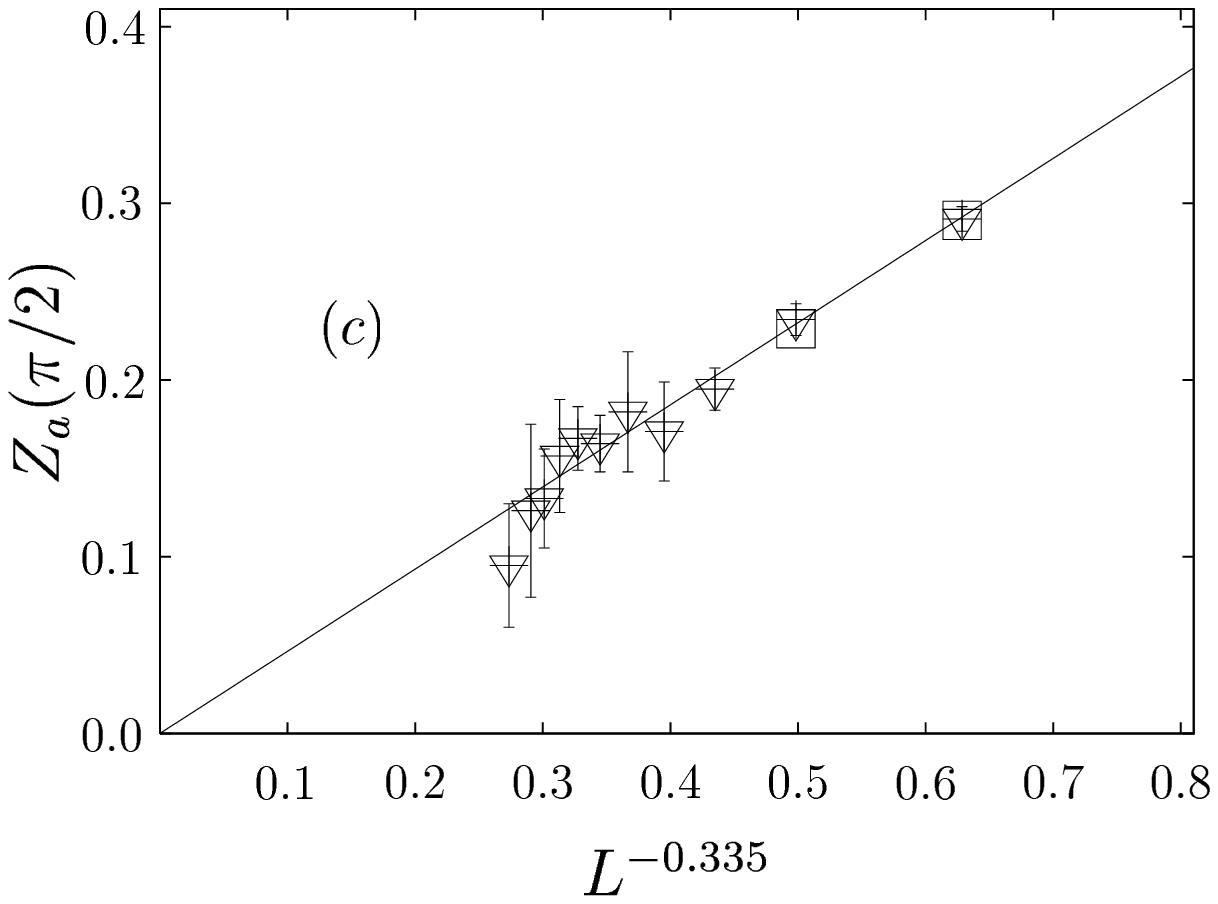,width=\gkl}\epsfig{file=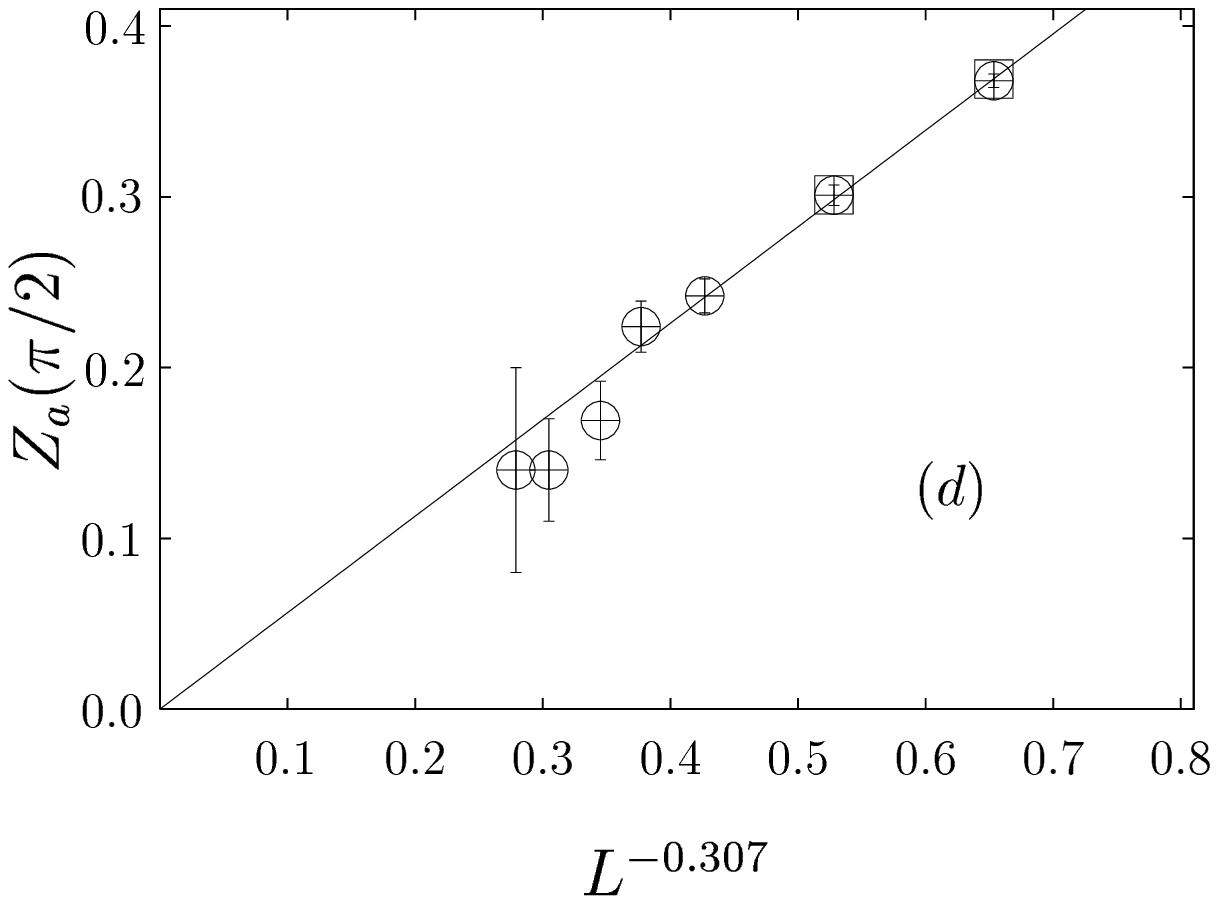,width=\gkl} 
\caption{Finite-size scaling for the quasiparticle weight   
at $ k_{||} =\pi/2$ for the bonding band ($k_\perp=0$) and its 
shadow ($k_\perp=\pi$) in the 2LL   
at $J/t=1.2$ (a), in the symmetric channel of the 3LL (b) at $J/t=1$ 
and $J/t=2$ [$k_{||}=3\pi/4$ for the bonding band (full symbols) and 
$k_{||}=\pi/4$ for the antibonding band (open symbols)],
and in the antisymmetric band of the 3LL at $k_{||}=\pi/2$
for $J/t=1$ (c), and for $J/t=2$ (d).   
The squares are results from ED. \label{fig:QP}} 
\end{figure}

We now address the QP weight  
$Z(k)=|\langle \Psi^{N-1}_0(k)|c^{}_{\downarrow}|\Psi^{N}_0\rangle|^2$  
which can be obtained from  
the imaginary time Green's function as the weight of  
the exponential with the slowest decay  
at large $\tau t$.  
As can be seen in Fig.~\ref{fig:QP},  
the QP weight is finite in the thermodynamic limit for both the bonding  
band and its shadow in the 2LL,  
and large compared to the 2D $t$-$J$ model~\cite{brunner00}   
(this result is valid for  
all $k$-points where the data is accurate enough to extract the QP weight).  
On the contrary, finite-size scaling of the QP weight in the antisymmetric   
channel of the 3LL leads to a vanishing QP weight in the thermodynamic limit,  
as expected for a Luttinger liquid. The scaling of $Z(k)$ can be fitted  
by $a*L^b$ with $a=0.464 \pm 0.014, 0.564\pm 0.015$ and   
$b=-0.335 \pm 0.015,-0.306\pm 0.016$ for $J/t=1,2$ respectively. Squares
in Fig.\ \ref{fig:QP} (c) and (d) correspond to ED results that confirm 
the QMC ones. 
The values obtained for the exponent $b$ depart from what would be obtained  
by na\"\i vely using the strong coupling result, that for $J=t$ leads to 
$t_{eff}=0.5$ (i.e.\ the supersymmetric case) where the exponent 
should be $b=-0.5$ \cite{sorella96}. 
In contrast to the results for the 
antisymmetric channel, the QP weight stays finite for  
the symmetric channel in the thermodynamic limit [Fig.\ \ref{fig:QP}(b)], 
although smaller than in the 2D case \cite{brunner00}. 
Therefore, our results demonstrate that the antisymmetric channel
in the 3LL 
{will evolve upon doping towards} 
the 
Luttinger-Tomonaga universality class, whereas the symmetric channel
belongs to the Luther-Emery one, as suggested previously \cite{rice97haas}. 

Summarizing, we have studied the spectral properties of a single hole in  
the $t$-$J$ model on 2LL (up to $2 \times 96$ sites) and 3LL   
(up to $3 \times 64$ sites) using a recently developed QMC algorithm.  
It is shown, that expansions around the limit $J_\perp \gg J_\parallel$  
describe accurately the bonding band and the high energy portion of the   
antibonding band in the 2LL, even in the isotropic case. However, such an   
expansion misses the low-energy portion of the antisymmetric channel, that   
corresponds to a shadow of the bonding band. For the 3LL strong  
coupling expansions ($J_\perp \gg J_\parallel$) for the antisymmetric channel  
and the 2LL give a good description of the antisymmetric and symmetric  
channels respectively, as long as $J_\perp$ is appreciably larger than   
$J_\parallel$. The dispersions in the isotropic 3LL  
seem to be less well described by a strong coupling expansion. The  
QP weight extrapolates in the thermodynamic limit  
to a large finite value for the bonding band and its shadow in the 2LL.  
For the 3LL finite-size scaling leads to a vanishing QP weight for   
the antisymmetric channel but a finite one for the symmetric channel,  
demonstrating that the first one corresponds to the Luttinger universality
class, whereas the latter corresponds to the Luther-Emery one.

We wish to thank D. Poilblanc, M.~Troyer and O.P.~Sushkov for helpful   
discussions.   
This work was supported by Sonderforschungsbereich 382  
and the Australian Research Council. 
We thank HLRS Stuttgart and IDRIS (Orsay) for allocation of 
CPU time on the CrayT3E and NEC SX5, respectively. 
$ $

\end{multicols}

\begin{thebibliography}{10}
\bibitem{fn1}
{This procedure was adopted in order to make smaller structures visible.
The integrated value of
the spectral function is normalized to $\pi/2$ for all
momenta $k$.
}
\end{thebibliography}
\end{document}